# Some analytic results in coherent quantum transport

Mohammad Mardaani, Keivan Esfarjani
Department of Physics, Sharif University of Technology, 11365-9161 Tehran, Iran

**Email:** mardaani@mehr.sharif.edu, k1@sharif.edu

## Abstract

A quantum wire of uniform cross section (but with eventual disorder) with three regions: dot, left lead, and right lead, is considered. Assuming that the same unitary transformation diagonalizes all unit cells of this wire, we propose a new formula for the calculation of the Green's function (GF) and the coherent transmission coefficient. This formula allows to calculate these quantitites much faster than the standard methods. In particular, the problem of a uniform dot (simple cubic uniform dot attached to the simple cubic wire), with all onsites equal and all hoppings equal is solved fully analytically. The energy and dot-length dependence of the GF, local density of states (LDOS), the transmission coefficient and bound state energies are also derived.

## Introduction

The study of electron transport in mesoscopic systems is one of the most fundamental problems in nanostructure physics; it has changed our understanding of transport phenomena in condensed matter systems. In recent years there has been a growing interest in electrical transport in quantum dots, nanowires, nanocrystals, and single molecules. Quantum wires and quantum dots are the fundamental building blocks for nanoelectronic devices. Some authors consider devices that consist of a dot connected to two continuum reservoirs of electrons (usually metallic leads) [1-15]. There has also been an increasing interest in the theoretical modeling of molecular wire systems.
Transport measurement is a powerful tool for the investigation of electronic properties of materials. Experimental research on molecular wires has increased over the past few years looking into the possibility of rectification and other phenomena [16-23].
Two different methods have been developed for the study of electron-impurity potential scattering, where by impurity potential, we mean any external inhomogeneity caused by impurities or any change in the shape of the walls etc, which make the Hamiltonian of the dot change under lattice translations. One method is based on a solution of the Lippmann-Schwinger equation and the other solves the **t** (*transmission*) matrix[5,6,7].
A fundamental idea behind these advances has been due to Landauer [13] who stated that conductance is proportional to transmittance. We rely on Landauer theory [15] as the basis for studying the conductance properties of quantum dot systems. This relates the lead to lead current to the transmission probability for an electron to scatter through the quantum dot.
In this paper, we study the transport properties in the ballistic regime of a cubic uniform quantum dot in the presence of metallic leads.
(The words lead and wire have been used interchangeably in this paper).

## Model

In this section, we present our model and assumptions, and derive an analytic formula allowing to calculate the conductance of a uniform dot of arbitrary size and shape analytically, and in the case of a disordered dot, we present an algorithm to calculate it in a much faster way. Actually The usual task of matrix inversion of the dot Hamiltonian is of the order of the cube of the number of orbitals (basis functions) in the dot. In this work, however, we show that under some assumptions, this task is reduced to the cube of the number of unit cells, the GF and the transmission coefficient of a one-band chain can be calculated analytically.

***<u>Assumptions of the model:</u>***

Both the dot and the leads are assumed to have the same geometry namely that of a cross section of arbitrary shape repeated periodically, this unit cell being the same for both leads and the dot. We further assume that all hopping integrals from one unit cell to the next are the same regardless of atom positions within the unit cell, but the onsite energies need not be the same within a cell, or from one unit cell to the next in the dot region. They need, however, to be only scaled from one cell to the next. Put in more mathematical terms, we require that the same unitary transformation diagonalizes any unit cell of the system. Based on these assumptions, we can write the three Hamiltonians in the left and the right leads and the dot in the following form:

$$H = \sum_{x,y,z} E_{x,y,z} \,| x, y, z >< x, y, z | + \sum_{x,y,x'y'z} V_{xyz,x'y'z} \left( | x, y, z >< x', y', z | + c.c \right) + \sum_{x,y,z} W_{xyz,xyz+1} \left( | x, y, z >< x, y, z+1 | + c.c \right) \quad (1)$$

where $z$ refers to unit cell number in the direction along the wire and $c.c$ refers to complex conjugate of $| x, y, z >< x', y', z |$ , and $x, y$ denote transverse degrees of freedom within the cell. $E$ is the onsite energy and $V$ refers to hopping integrals. The first hopping integral in the above formula corresponds to intracell, and the second one to intercell hoppings. It is assumed that intercell hoppings are all the same $V_{xyz,xyz+1} = v_{z,z+1}$. One can consider a single cell Hamiltonian and rewrite it in a basis where it is diagonal.

$$H = \sum_{z} \left[ H_{cell}(z) + V(z, z+1) + c.c \right] \quad (2)$$

where $H_{cell}(z) = \sum_{t} \varepsilon_{tz} |tz\rangle\langle tz| \quad and \quad V(z, z+1) = \sum_{t} v_{t,z,z+1} |t, z\rangle\langle t, z+1|$

Here, $t = 1, ..., p$ labels the diagonalized transverse degrees of freedom obtained from the original $x, y$ basis by a unitary transformation so that the first two terms in equation (1) are transformed into $H_{cell}$. Furthermore, the cell has been chosen so that electron hopping only occurs between nearest neighbor cells.

Thus, under our assumptions, the system of "thick" wires can be seen as $p$ independent one-band chain, where most calculations can be done analytically. This assumption is valid for the case of most quantum wires such as carbon nanotubes. We can also see here, that the general task of inverting a $N_xN_yN_z$ matrix (which is of order $(N_xN_yN_z)^3$ ) can, even in the case of a disordered system, be reduced to that of inverting $N_xN_y$ times a matrix of size $N_z$ only. This latter task is of order $(N_xN_yN_z)^3$ if standard matrix inversion algorithms are used. For unidimensional systems, however, faster algorithms using the band property of the Hamiltonian matrix can make this task even faster than $O(N^3)$.

Consequently, in what follows, we will assume a one-band chain composed of a central region (dot) and two leads, and will derive the analytic expression for the GF and the transmission coefficient assuming the dot is uniform (no disorder). There will be three onsite energies and hopping integrals, one in each region, plus two hoppings from the dot to the two leads.

For a general system divided into two parts $A$ and $B$, it is easy to show that the total ( $A+B$ ) GF projected onto the Hilbert space of the dot (called $A$ below), can be written as:

$$G_A^{-1}(z) = G_{A0}^{-1}(z) - H_{AB}G_{B0}(z)H_{BA} = z - H_A - H_{AB}G_{B0}(z)H_{BA} \quad (3)$$

where the index 0 refers to the GF of the isolated subsystem ( $A$ or $B$ ). The second term on the right is called the self-energy and is the effect of connecting the subsystem $B$ to $A$, and can also be defined in case of many subsystems $C$, $D$ etc…
connected to $A$. It is called self-energy because it is the term added to the Hamiltonian of subsystem $A$, so that it modifies the energy eigenvalues of the system. The self-energy has nonzero matrix elements on the sites directly connected to $B$, and is therefore local. For a linear dot of size N, it has nonzero matrix elements only on the two end sites 1 and $N$. In the complex number $z = E \pm i\eta$, + is chosen for the retarded GF, and – for the advanced GF.

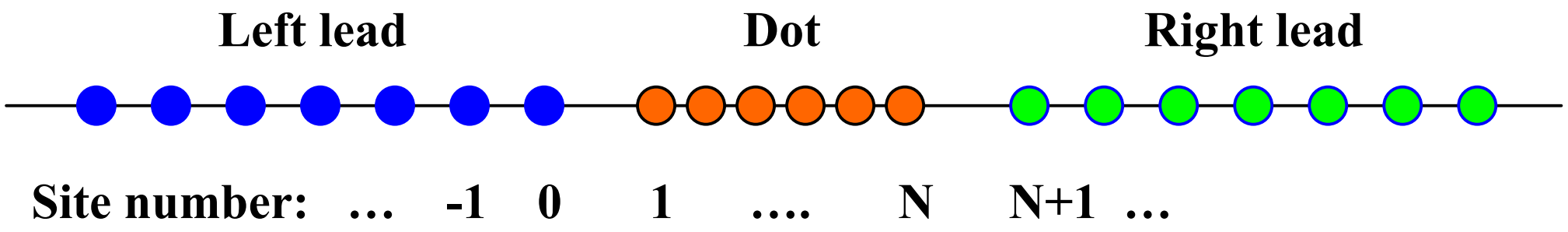

**FIG1**.quantum dot attached to the two various leads

Below, we will compute it analytically for a linear dot attached to two leads denoted by left and right, and the expression for the total GF of the dot will be deduced. for the transverse mode $(m,n)$, the diagonal parts of the GF on sites 1 and $N$ N are modified as follows:

$$[G_{mn}^{-1}]_{11} = [G_{0,mn}^{-1}]_{11} - \Sigma_{L,mn}; \qquad \Sigma_{L,mn} = \frac{t_{D,L}^2}{t_L} e^{-i|k_{L,mn}|a} \qquad (4.1)$$

$$[G_{mn}^{-1}]_{NN} = [G_{0,mn}^{-1}]_{NN} - \Sigma_{mn,R}; \qquad \Sigma_{mn,R} = \frac{t_{D,R}^2}{t_R} e^{-i|k_{R,mn}|a} \quad (4.2)$$

where $a$ refers to the lattice constant.

since the GF of the isolated lead $\alpha = L, R$ projected on its last atom, neighbor to the dot is:

$$t_\alpha G_\alpha(z) = [(z - E_\alpha)/2t_\alpha] \pm \sqrt{[(z - E_\alpha)/2t_\alpha]^2 - 1} = e^{-i|k_\alpha|a} \quad (5)$$

where $z = E_\alpha + 2t_\alpha \cos(k_\alpha a)$

Here $t_\alpha, E_\alpha$ are respectively the hopping and the onsite energy within the lead $\alpha$. The sign before the square root is chosen such that the imaginary part of the retarded GF becomes negative (in order to obtain a positive density of states). This gives the – sign in the exponent $-i|k|a$, and for the advanced function one needs to take $+i|k|a$ in the exponent. Note that the complex number $k_\alpha$ becomes a function of $z$. Thus the self-energy due to the lead connections is only added to the left and right diagonal elements of the Hamiltonian of the dot. The dot-lead contact does not change, or changes negligibly, the spectrum of the leads since it affects only one of their sites and the leads are infinite. All other matrix elements of the inverse GF of the dot are unmodified by the leads:

$$[G_{mn}^{-1}]_{ij} = [E + i\eta - E_{mn,i} - \Sigma_{L,mn}\delta_{i,1} - \Sigma_{R,mn}\delta_{i,N}]\delta_{ij} - t_{D,i}(\delta_{i,j+1} + \delta_{i,j-1}) \quad \text{for } i, j = 1, \ldots, N \quad (6)$$

The transmittance is related to the GF by the Landauer's formula:

$$T(E) = tr(\Gamma_L G^{(+)} \Gamma_R G^{(-)}) \quad (7)$$

In this formula, $\Gamma_\alpha$ is -2 times the imaginary part of the self-energy of the lead $\alpha$, and for a single transverse channel $(m,n)$, the transition rate $\Gamma_\alpha$ to or from the lead $\alpha$ can be simplified as:

$$\Gamma_\alpha = -2\,\mathrm{Im}\,\Sigma_\alpha = 2\frac{t_{\alpha-Dot}^2}{t_\alpha}\sin|k_{\alpha,mn}|a = \frac{t_{\alpha-Dot}^2}{t_\alpha^2}\frac{|v_{\alpha,mn}|}{a} \quad (8)$$

where $v_\alpha$ is the group velocity of the channel $(m,n)$ in the lead $\alpha$, and $G^{(+)}, G^{(-)}$ are the advanced and retarded GFs of the dot respectively, and we have assumed $\hbar = 1$. In terms of the complex energy variable $z$, the single channel transition rate can also be written as:

$$\Gamma_{\alpha,mn}(z) = 2\frac{t_{D,\alpha}^2}{t_\alpha}\sqrt{1 - \left(\frac{z - E_{\alpha,mn}}{2t_\alpha}\right)^2} \quad (9)$$

If one goes back to its definition, one can clearly see that $\Gamma_\alpha$ is zero outside the energy band of the lead $\alpha$ which is $[E_{\alpha,mn} - 2t_{\alpha,mn}, E_{\alpha,mn} + 2t_{\alpha,mn}]$. Furthermore, only the matrix element 11 and NN are nonzero for $\alpha$ =L and R, respectively.

***Analytical results for a homogeneous dot:***

In the case where the dot has uniform onsite and hoppings (no disorder), the chain problem, and hence the full quantum wire system can be solved analytically. The inverse GF of the dot differs from its isolated value only by the two self-energies added to the sites 1 and N. In this case, we show that it can be inverted analytically. For calculation of the density of states (DOS) only the diagonal elements are needed, and for the conductance only $G_{1N}$ is needed.

After dividing by the hopping integral, the inverse of the GF of the dot has the following form:

$$M(N,\alpha_{mn},\beta_{mn}) = \begin{bmatrix} x_{mn}-\alpha_{mn} & -1 & 0 & \ldots & 0 & 0 & 0 \\ -1 & x_{mn} & -1 & \ldots & 0 & 0 & 0 \\ 0 & -1 & x_{mn} & \ldots & 0 & 0 & 0 \\ \ldots & \ldots & \ldots & \ldots & \ldots & \ldots & \ldots \\ 0 & 0 & 0 & \ldots & x_{mn} & -1 & 0 \\ 0 & 0 & 0 & \ldots & -1 & x_{mn} & -1 \\ 0 & 0 & 0 & \ldots & 0 & -1 & x_{mn}-\beta_{mn} \end{bmatrix}_{(N\times N)} \qquad (10)$$

$$D(N,\alpha_{mn},\beta_{mn}) \equiv \mathrm{Det}[M(N,\alpha_{mn},\beta_{mn})]$$

$$D_p \equiv D(p,0,0)$$

where $x_{mn} = \dfrac{z - E_{Dot,mn}}{t_{Dot,mn}}$; $\alpha_{mn} = \dfrac{\Sigma_{mn,L}(z)}{t_{Dot}}$; $\beta_{mn} = \dfrac{\Sigma_{mn,R}(z)}{t_{Dot}}$

From now on, the channel number $(m,n)$ will be dropped for simplicity since we restrict ourselves to a simple one-orbital one-atom chain.

This determinant is calculated in the Appendix (for uniform dot). For different GFs one finds in units of $1/t_{Dot}$ that (for a single mode the index $(m,n)$ is removed):

$$G_{1N} = \frac{1}{D(N,\alpha,\beta)}; G_{11} = \frac{D(N-1,0,\beta)}{D(N,\alpha,\beta)}; G_{NN} = \frac{D(N-1,\alpha,0)}{D(N,\alpha,\beta)};$$
$$G_{ii} = \frac{D(i-1,\alpha,0)D(N-i,0,\beta)}{D(N,\alpha,\beta)} \qquad (11)$$

The transmission coefficient $T(E)$ of a single chain (mode) can be written in the following form:

$$G = \frac{2e^2}{h}T(E)$$

$$T(E) = \Gamma_R\Gamma_L\,|\,G_{1N}\,|^2 = \frac{v_L\, v_R}{a^2}\left|\frac{t_{D,L}\, t_{D,R}}{t_D^2 t_L t_R}\frac{1}{D(N,\alpha,\beta)}\right|^2 \qquad (12)$$

Thus the calculation of $D(N,\alpha,\beta)$ is necessary as all the above quantities are expressed as a function of $D$. We find that it can be written as follows (see Appendix):

$$D(N,\alpha,\beta)=\begin{cases}\dfrac{\sin(N+1)\Phi}{\sin\Phi}-(\alpha+\beta)\dfrac{\sin N\Phi}{\sin\Phi}+\alpha\beta\dfrac{\sin(N-1)\Phi}{\sin\Phi}\\ N[(1-\alpha)(1-\beta)]+1-\alpha\beta\\ \dfrac{\sinh(N+1)\Phi}{\sinh\Phi}-(\alpha+\beta)\dfrac{\sinh N\Phi}{\sinh\Phi}+\alpha\beta\dfrac{\sinh(N-1)\Phi}{\sinh\Phi}\end{cases} \quad (13)$$

where $\frac{x}{2}=\frac{z-E_D}{2t_D}=\begin{cases}\cos\Phi & \text{if } |x|<2\\ \pm 1 & \text{if } |x|=2\\ \cosh\Phi & \text{if } |x|>2\end{cases}$

# Results:

## ***Dependence on dot length and energy***

For energy ranges *outside the dot band,* it can easily be seen that the behavior of $G_{1N}$ as a function of the energy is exponential since $\Phi$ becomes pure imaginary. A decay length may be obtained in this limit. In units of the lattice constant, it is the inverse of the coefficient of $N$ in $\log[G_{1N}]$:

$$G_{1N}\longrightarrow\exp\left(-\frac{(N-1)a}{\lambda(E)}\right)\quad \text{if}\quad |x|>2 \qquad (14)$$

With $\lambda(E)=a/\log[|\frac{E-E_D}{2t_D}|+\sqrt{(\frac{E-E_D}{2t_D})^2-1}]$.

Likewise, the transmission coefficient T will have a similar behavior since it is quadratic in $G_{1N}$.

Within the dot band, however, the behavior of $G_{1N}$ will be oscillatory with respect to $N$. Finally, if the energy lies on the border of the band, to leading order, $G_{1N}$ becomes proportional to $1/N$, and the transmission coefficient proportional to $1/N^2$ !

The exponential length-dependence outside the band region was previously known[3], although the authors do not know of any analytic proof of it. It has also been numerically checked by Joachim and Vinuesa [7].

*At the center of the dot band*, $x=0$ and $\Phi=\pi/2$. The calculations will become simple if onsite energies of the leads are the same as the dot's. The self-energies ($\alpha$ and $\beta$) become pure imaginary: $\alpha=\frac{t_{D,L}^2}{t_Dt_L}i;\beta=\frac{t_{D,R}^2}{t_Dt_R}i$. Results will depend on the parity of $N$, but independent of dot size.

Onsite GF matrix elements are then a pure imaginary number. Within a factor of -i$\pi$, $G_{jj}$ is the LDOS on each atom except for the atoms at the two ends where either $\alpha$ or $\beta$ should be set to zero in the above expressions. One obtains for odd lengths:

$$N=2n;\ T=\frac{4|\alpha\beta|}{(1+|\alpha\beta|)^2};\ G_{jj}(0)=\begin{cases}\dfrac{\alpha}{\alpha\beta-1} & j=2k\\ \dfrac{\beta}{\alpha\beta-1} & j=2k+1\end{cases} \qquad (15)$$

$$N=2n+1; T=\frac{4|\alpha\beta|}{|\alpha+\beta|^2};\ G_{jj}(0)=\begin{cases}\dfrac{\alpha}{\alpha+\beta} & j=2k\\ \dfrac{\beta}{\alpha+\beta} & j=2k+1\end{cases}$$

We see that the LDOS(*E=0*) alternates on all sites and its value does not depend on N; furthermore, the diagonal elements of the GF are pure imaginary at that energy.

***<u>Bound states in the uniform dot attached to leads.</u>***

There can be formation of bound states in the dot if there is an energy range within the bandwidth of the isolated dot which falls below or above all the energy levels of the leads. The bound state energies are the roots of the following equation (note that $\alpha$ and $\beta$ are functions of the energy $E$).

$$F(E)=t_{Dot}G_{1N}\sin\Phi(E)=\sin(N+1)\Phi-(\alpha+\beta)\sin N\Phi+\alpha\beta\sin(N-1)\Phi=0, \qquad (16)$$

In FIG. 2, the function F has been displayed as a function the energy for a system where all onsite energies are equal to 0 and the hopping of the dot is twice that of the leads. The four curves correspond to 4 different values of dot-to-lead hoppings. The bound state energies are the intersection of the curve of *F(E)* with the x axis. The curve of *F* is even in energy and the same bound states also exist in the [-2,-1] *eV* energy range.

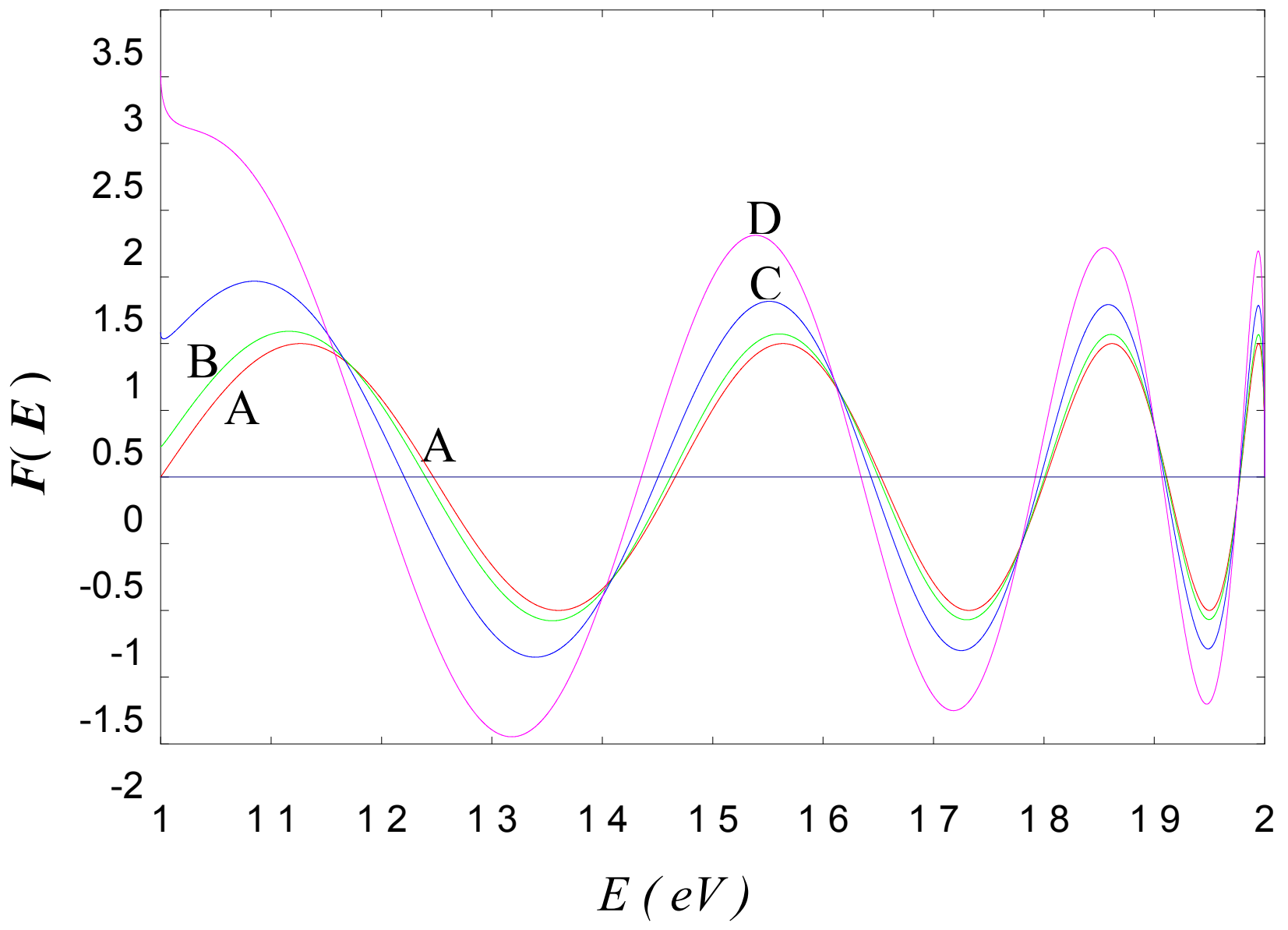

**FIG 2**: Bound state energies of a 20 atom dot attached to two leads of same onsite

4

energy with twice smaller hopping integral. The four curves are obtained for different dot-lead hoppings:

$$F(E);\quad N=20; E_{0,D}=E_{0,lead}=0\quad t_D=2t_{lead}=-1\ eV$$

$$\mathrm{t}_{\mathrm{D,lead}}=\{0.0(\mathrm{A:red}),0.25(\mathrm{B:green}),0.5(\mathrm{C:blue}),0.75\ \ (\mathrm{D:pink})\}$$

Below, the bound states are found analytically in the limit of small dot-lead hopping ($t_{D,L} << t_D$ and $t_{D,R} << t_D$). To find the bound state energies, we must solve the following nonlinear equation:

$$\begin{cases} E = 2t_D \cos(\dfrac{\pi j + \delta}{N+1}) \\ \tan(\Phi - \delta) = \dfrac{(1-\alpha\beta)\sin\Phi}{(1+\alpha\beta)\cos\Phi - \alpha - \beta} \end{cases} \tag{17}$$

where $\Phi = \cos^{-1}(\dfrac{E}{2t_D})$

For isolated $N$ Atom's dot we know:

$$F(E) = \prod_{j=1}^{N} \left( E - 2t_D \cos(\frac{j\pi}{N+1}) \right) \tag{18}$$

for weak lead – dot interaction ($\alpha, \beta << 1$ and $\alpha\beta \approx 0$) we have the following result:

$$F(E) = \prod_{j=j_{\min}}^{N} \left( E - 2t_D \cos(\frac{j\pi}{N+1}) - \Delta_j \right) = 0 \quad \Rightarrow E_{j,BS} = 2t_D \cos(\frac{j\pi}{N+1}) + \Delta_j \tag{19}$$

where

$$\Delta_j = -2t_D(\alpha_j + \beta_j)\sin^2(\frac{\pi j}{N+1}) \;\; ; \;\; \alpha_j = \frac{t_{D,L}^2}{t_D t_L}\left( |\frac{t_D}{t_L}\cos(\frac{\pi j}{N+1})| - \sqrt{\frac{t_D^2}{t_L^2}\cos^2(\frac{\pi j}{N+1}) - 1} \right)$$

$\beta_j$ is obtained from $\alpha_j$ by changing $(L \to R)$

***<u>Results for a nanocrystal with cubic structure:</u>***

In the instance of a wire with cubic structure, the transverse modes can be found analytically. The Hamiltonian of this system in the tight binding basis and in the presence of an onsite potential after diagonalization can be written as follows:

$$H = \sum_{m,n,z} [\ E_{mn,z} C^{\dagger}_{mn,z} C_{mn,z} - t_{mn,z,z+1}(C^{\dagger}_{mn,z} C_{mn,z+1} + C^{\dagger}_{mn,z+1} C_{mn,z})\ ] \tag{20}$$

where $E_{mn,z}$ and $t_{z,z+1}$ are respectively the onsite energies and hopping integrals in the $(m,n)$ mode, $C^{\dagger}_{mn,z}$ and $C_{mn,z}$ are electron creation and annihilation operators on longitudinal site ( $z$ ) of a quasi 1-D chain with lattice constant $a$. The onsite and hopping energies, after diagonalizing the modes of a unit cell, are (here $L$ refers to the leads assumed identical):

$$E_{mn,z} = \begin{cases} E_{mn,D} = E_0 + 2t_{D,x}Cos(\frac{m\pi}{N_x+1}) + 2t_{D,y}Cos(\frac{n\pi}{N_y+1}) & z = 1,\ldots,N \\ E_{mn,L} = 2t_{L,x}Cos(\frac{m\pi}{N_x+1}) + 2t_{L,y}Cos(\frac{n\pi}{N_y+1}) & z < 1 \text{ or } z > N \end{cases} \qquad (21)$$

where

$$t_{mn,z,z+1} = \begin{cases} t_L & z < 0 \;\; or \;\; z > N \\ t_D & z = 1,\ldots,N-1 \\ t_{D,L} & z = 0, N \end{cases}$$

for simple cubic dot attached to simple cubic wire, DOS(E) are found to have the following form:

$$\text{DOS}(E) = -\frac{1}{\pi t_{Dot}} \text{Im} \sum_{mn,j} \left( \frac{D(j-1,\alpha_{mn},0)D(N-j,0,\beta_{mn})}{D(N,\alpha_{mn},\beta_{mn})} \right) \qquad (22)$$

in the following table for a $(m,n)$ mode, the transmission coefficient has been calculated in the different ranges of the Fermi energy.

**Table.1**: transmission coefficient for a $(m,n)$ mode in the different energy range

| **Energy range** | $T_{mn}(E)$ |
|---|---|
| Inside wire and dot E-band | $\frac{4\kappa^2\sin^2(\theta)\sin^2(\phi)}{[\kappa^2\cos(2\theta)\sin(N-1)\phi-2\kappa\cos(\theta)\sin N\phi+\sin(N+1)\phi]^2+[\kappa^2\sin(2\theta)\sin(N-1)\phi-2\kappa\sin(\theta)\sin N\phi]^2}$ |
| Inside wire E-band ,but lies on the border of the dot E-band | $\frac{4\kappa^2\sin^2(\theta)}{[\kappa^2(N-1)\cos(2\theta)-2\kappa N\cos(\theta)+N+1]^2+[\kappa^2(N-1)\sin(2\theta)-2\kappa N\sin(\theta)]^2}$ |
| Inside wire E-band, but outside dot E-band | $\frac{4\kappa^2\sin^2(\theta)\sinh^2(\phi)}{[\kappa^2\cos(2\theta)\sinh(N-1)\phi-2\kappa\cos(\theta)\sinh N\phi+\sinh(N+1)\phi]^2+[\kappa^2\sin(2\theta)\sinh(N-1)\phi-2\kappa\sin(\theta)\sinh N\phi]^2}$ |
| Outside wire and dot E-band | 0 |

$$\text{Where: } \kappa=\frac{t_{L,D}^2}{t_L t_D}\ ;\ \theta=\cos^{-1}\left(\frac{E-E_{mn,L}}{2t_L}\right)\ ;\ \phi=\cos^{-1}\left(\frac{E-E_0-E_{mn,Dot}}{2t_{Dot}}\right)$$

(here $L$ refers to the leads assumed identical):

if $\theta=\phi=0$ then :

$$T=\frac{4\kappa^2}{(\kappa-1)^2[N(1-\kappa)+\kappa+1)]^2+4\kappa^2[N(\kappa-1)-\kappa]^2} \quad (23)$$

In **FIG.3** and **FIG.4**, DOS($E$) and $T(E)$ have been plotted for an ideal nanowire with square cross section; for simplicity all onsite and hopping terms (dot and lead) have been set respectively to 0 and 1 $eV$. The range of energy band of both wire and dot is [-6,6] $eV$. The total number of channels depends on number of transverse orbitals which is equal to 100 in the following case.

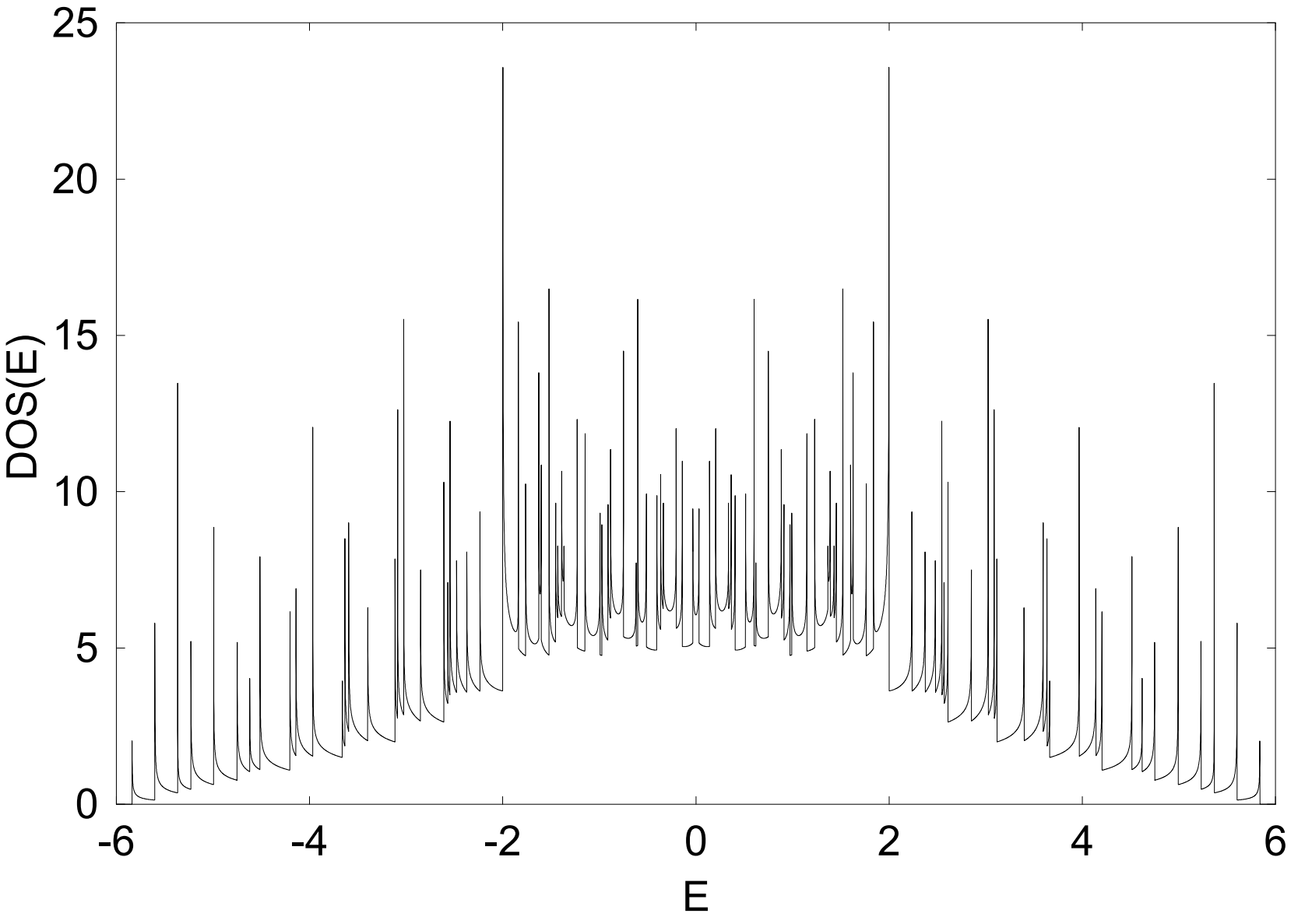

**FIG 3 :** this figure shows the Density of States DOS(E) vs. E (eV)[Fermi energy] for the following specific parameter(ideal wire):

$$N_x = N_y = 10; \quad t_D = t_{lead} = t_{D,lead} = 1 \ eV;$$

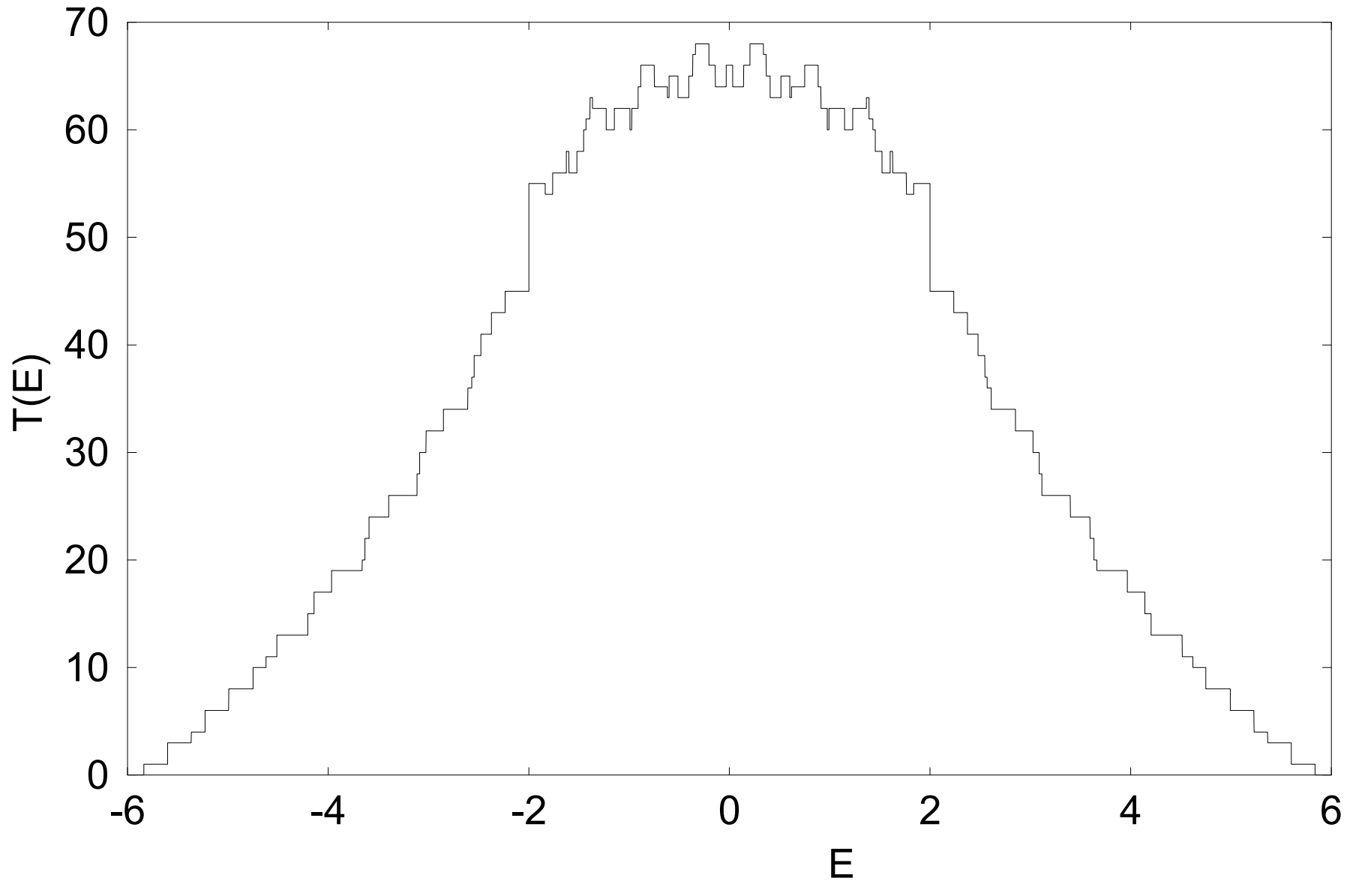

**FIG 4 :** this figure shows the transmission coefficient T(E) vs. E (eV)[Fermi energy] for the following specific parameters (ideal wire):

$$N_x = N_y = 10; \quad t_D = t_{lead} = t_{D,lead} = 1 \ eV;$$

In **FIG.5** and **FIG.6** log[DOS(E)] and log[T(E)] have been plotted for a simple cubic dot of three different lengths, attached to same right and left leads (with square cross section). all onsite energy of atoms in this system have been set to zero; but all hoppings in the dot have been set to half of hopping terms of the leads. The dot to lead hopping is one quarter of the hopping within the leads. The energy range of the leads and the dot is respectively [-3,3] $eV$ and [-6,6] $eV$. Within the dot band, in the [-3,3] $eV$ range, log [DOS (E)] has an oscillating behavior. But outside of this range it varies linearly as a function of E; this behavior being independent of the dot size.

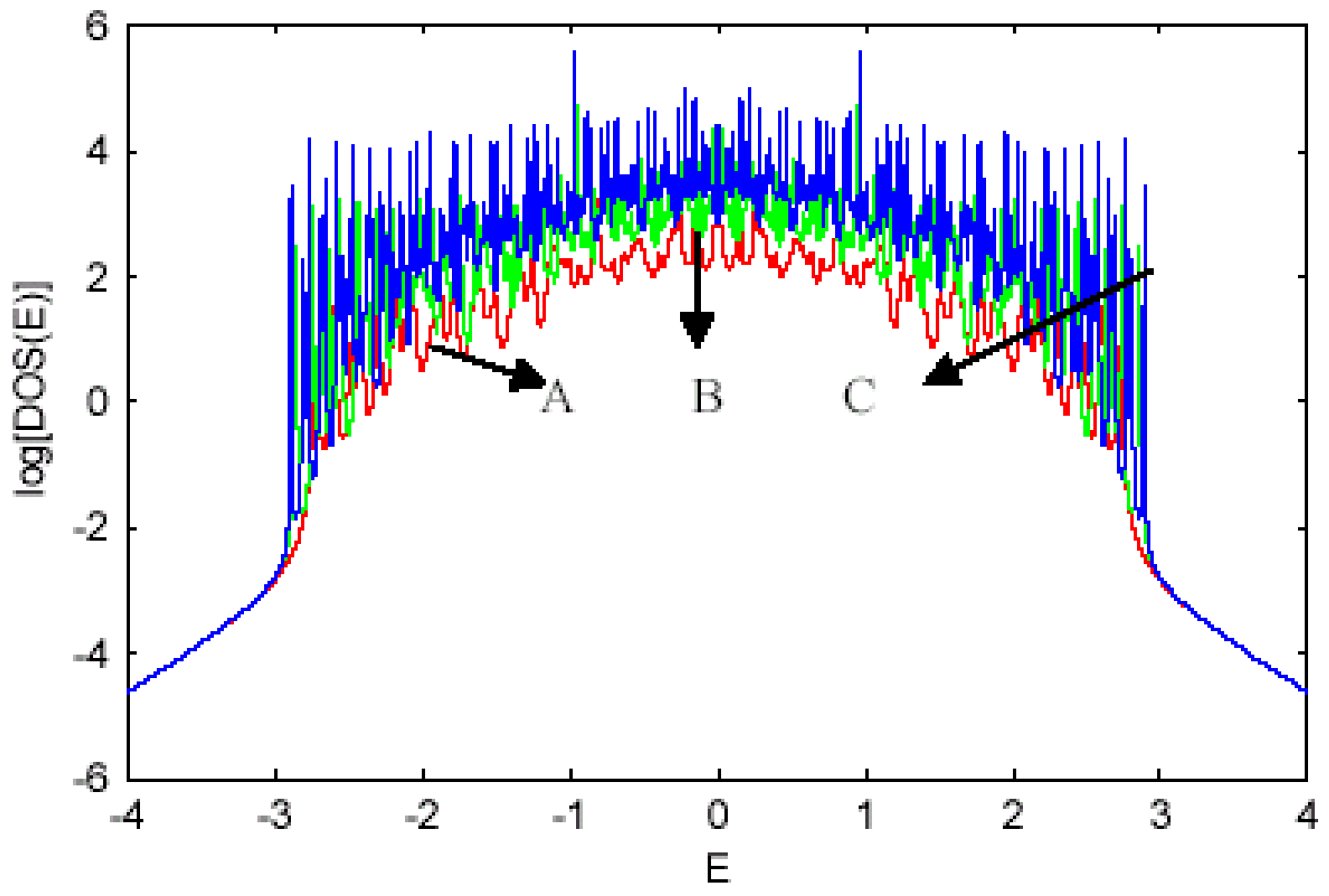

**FIG 5 :** this figure shows log[DOS(E)] vs. E (eV)[Fermi energy] for the following specific parameter:

$\log[\mathrm{DOS}(E)]; N_x = N_y = 10; N_z = 4n \;\; ; n = 1:(A), 2:(B), 3:(C)$

$t_{D,x} = t_{D,y} = t_{D,z} = \frac{1}{2} t_{lead} = 0.5 \; eV; \;\; t_{D,R} = t_{D,L} = \frac{1}{4} t_{lead} = 1 \; eV$

The exponential decay of the transmission, however increases with the dot length as can be seen from FIG. 7. The system being a metal (no gap within the band), the transmission never decays to zero inside the band.

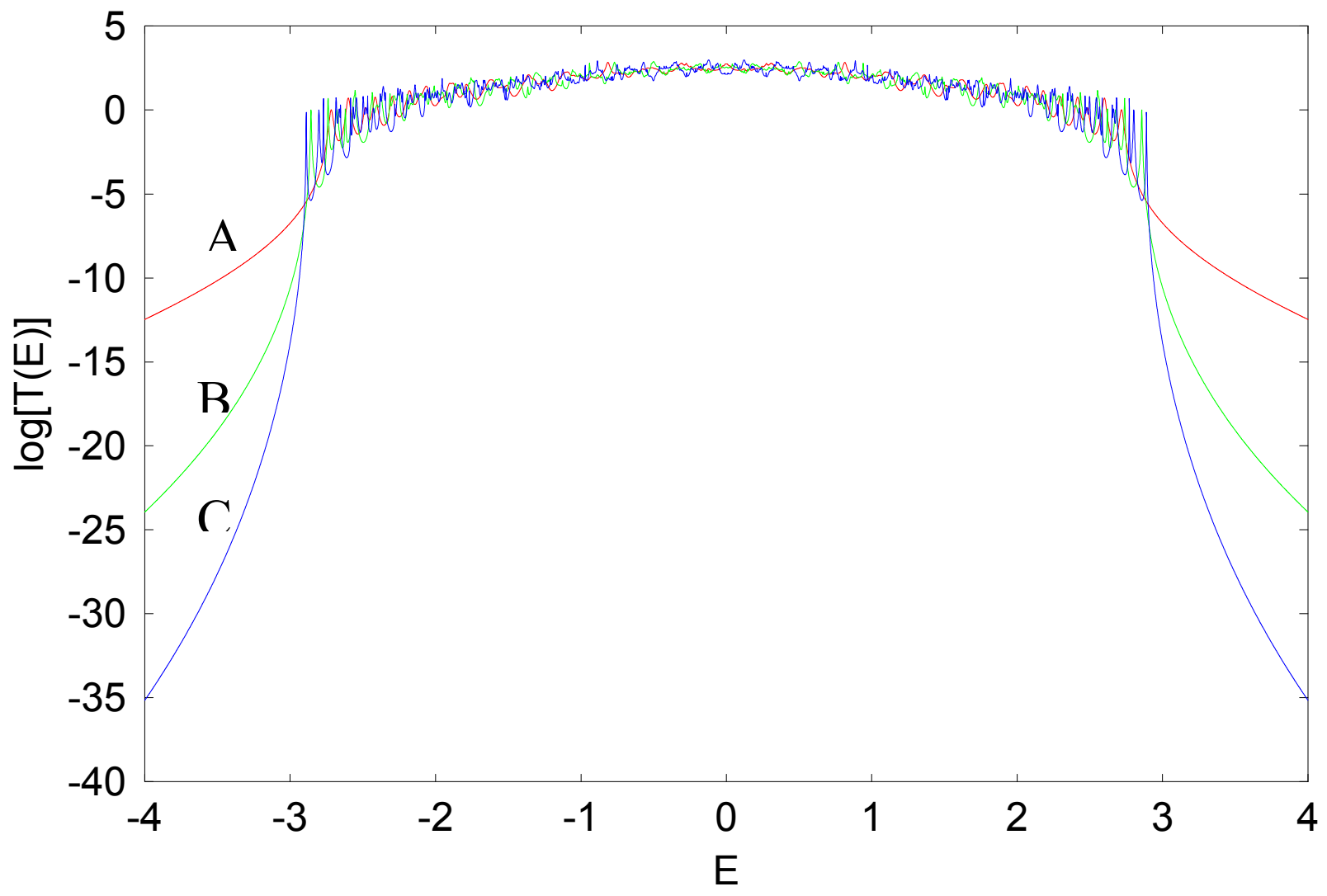

**FIG 6:** this figure shows log[T(E)] vs. E (eV)[Fermi energy] for the following specific parameter:

$$\log[T(E)]; N_x = N_y = 10; N_z = 4n \;\; ; n = 1:(A), 2:(B), 3:(C)$$

$$t_{D,x} = t_{D,y} = t_{D,z} = \tfrac{1}{2} t_{lead} = 0.5 \; eV; \;\; t_{D,R} = t_{D,L} = \tfrac{1}{4} t_{lead} = 1 \; eV$$

In the following graph, a uniform dot in which there is no transverse hopping was assumed. This could model a series of parallel polymer chains bridged between the two leads.

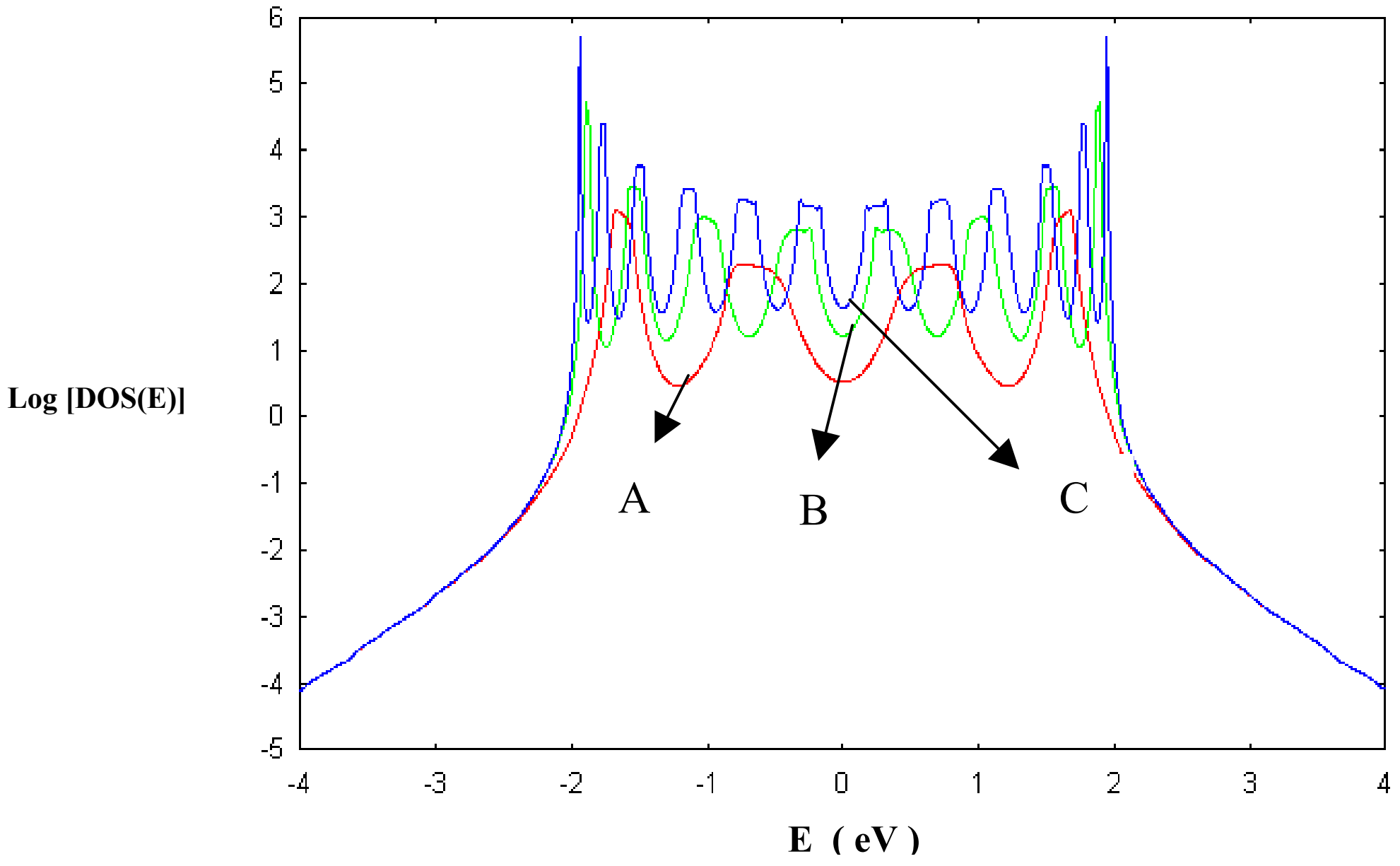

**FIG 7:** this figure shows $\log[DOS(E)]$ vs. E (eV)[Fermi energy] for the following specific parameter:

$$\log[Dos(E)]; N_x = N_y = 10; N_z = 4n; \;\; n = 1:(A), 2:(B), 3:(C)$$

$$t_{D,x} = t_{D,y} = 0; \;\; t_{D,z} = t_{lead} = 1 \; eV; \;\; t_{D,R} = t_{D,L} = 0.5 \; eV$$

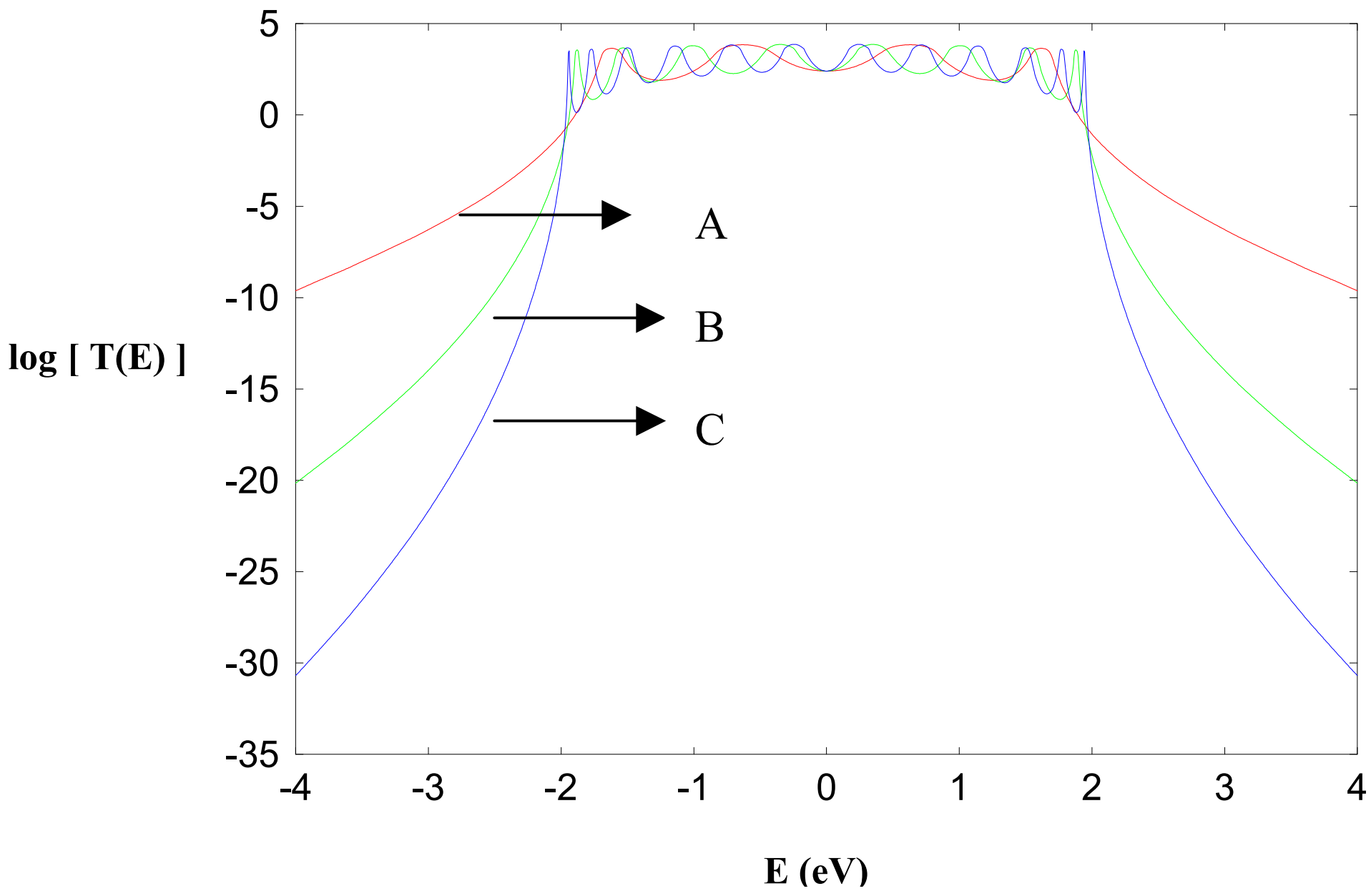

**FIG 8:** this figure shows $\log[T(E)]$ vs. E (eV)[Fermi energy] for the following specific parameter:

$\log[T(E)]; N_x = N_y = 10; N_z = 4n, \quad n = 1:(A), 2:(B), 3:(C)$

$t_{D,x} = t_{D,y} = 0; \quad t_{D,z} = t_{lead} = 1\ eV) t_{D,R} = t_{D,L} = 0.5\ eV$

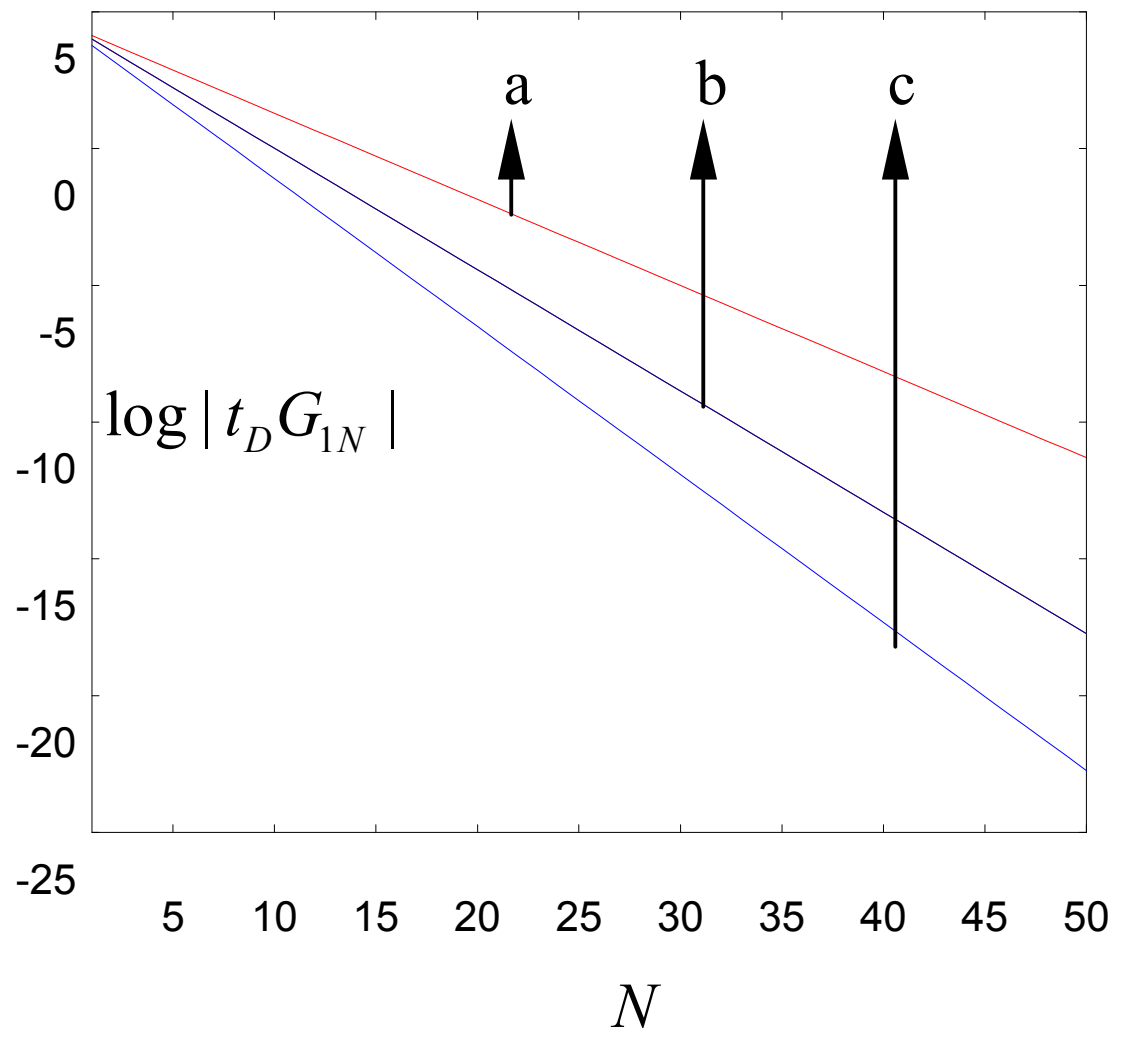

**FIG 9:** this figure shows $\log|t_D G_{1N}|$ vs. $N$ (dot size) for the following specific parameter:

$\log[G_{1N}(E,N)]; \quad N_x = N_y = 10; \quad t_{Dot,x} = t_{Dot,y} = 0$

$t_{Dot,z} = 1(eV); \quad E(eV) = 2.1:(a), \quad 2.2:(b), \quad 2.3:(c)$

If the Fermi energy lies inside the wire energy band but outside the dot energy band, ( $N >> 1$ or $E >> E_{mn}$ ) we have:

$$T(E) = \sum_{mn} \frac{16\kappa^2 \sin^2(\theta)\sinh^2(\phi)e^{-2(N-1)\phi}}{[e^{2\phi} + \kappa^2\cos(2\theta) - 2\kappa\cos(\theta)e^{\phi}]^2 + [\kappa^2\sin(2\theta) - 2\kappa\sin(\theta)e^{\phi}]^2} \quad (24)$$

The results can be summarized in the following table:

**Table2:** Behavior of $\log[T(E)]$ and $\log[\mathrm{DOS}(E)]$ vs. Fermi energy in the inside and outside of the dot energy band

| $E$ | $\lvert E\rvert < 2\lvert t_{D,x} + t_{D,y} + t_{D,z}\rvert$ | $\lvert E\rvert > 2\lvert t_{D,x} + t_{D,y} + t_{D,z}\rvert$ |
|---|---|---|
| $\log[T(E)]$ | Oscillating behavior | $-\beta(E)(N-1)$ |
| $\log[\mathrm{DOS}(E)]$ | Oscillating behavior | $\pm E$ |

## Conclusion

In this work, transmission coefficient and density of states of a uniform dot attached to two leads of the same transversal symmetry were calculated fully analytically. In general, it was shown that if all cross sections within the dot and the leads follow the same symmetry properties, i.e. each cross section can be diagonalized by the same unitary transformation, it is possible to reduce the task of calculation of the Green's function of the dot to linear in the transverse degrees of freedom: the problem is in fact reduced to $N_{transverse}$ strictly one-dimensional problems.

It was shown that within the dot and the wire energy ranges, the transmission and DOS have an oscillating behavior. At the band edge of the dot, they behave as $1/N^2$, and outside the dot band, the transmission decays exponentially with length, but the DOS is decays independently of the length. In the tunneling regime, where the Fermi energy lies outside the dot band, one can make a switch by applying a gate voltage to the dot. There is no flow of current at zero gate voltage, but if the latter can bring the Fermi energy within the dot band, the transmission will not be exponentially small anymore, and current will flow. Most quantum switches work on this basis, namely in systems which either display a band gap or have a different energy band width than the attached leads.

## Appendix

To calculate the inverse of $M$ , we first need its determinant:
$D(N,\alpha,\beta) = D_N - (\alpha+\beta) D_{N-1} + \alpha\beta D_{N-2}$ (A.1)
where $D_N$ is the determinant of the following matrix denoted by $M_N$ :

$$M_N = \begin{bmatrix} x & -1 & 0 & \ldots & 0 & 0 & 0 \\ -1 & x & -1 & \ldots & 0 & 0 & 0 \\ 0 & -1 & x & \ldots & 0 & 0 & 0 \\ \ldots & \ldots & \ldots & \ldots & \ldots & \ldots & \ldots \\ 0 & 0 & 0 & \ldots & x & -1 & 0 \\ 0 & 0 & 0 & \ldots & -1 & x & -1 \\ 0 & 0 & 0 & \ldots & 0 & -1 & x \end{bmatrix}_{(N\times N)} \qquad \text{(A.2)}$$

In calculating matrix elements of $G$ , the cofactors which appear in the numerator are also of the form of $M$ where either one of $\alpha$ or $\beta$ is replaced by zero. It is easy to show that $D_N$ satisfies the following recursion relation:
$D_N = xD_{N-1} - D_{N-2}$; with $D_0 = 1$ and $D_1 = x$ .
The solution results in:

$$D_N = \prod_{j=1}^{N} \left[ x + 2Cos(\frac{\pi j}{N+1}) \right] = \begin{cases} \dfrac{\sin\,(N+1)\Phi}{\sin\Phi} & \text{if } \; |x| < 2 \\ N+1 & \text{if } \; |x| = 2 \\ \dfrac{\sinh\,(N+1)\Phi}{\sinh\Phi} & \text{if } \; |x| > 2 \end{cases} \qquad \text{(A.3)}$$

for an isolated quantum dot:

$$G_{ij} = [M^{-1}]_{ij} = [M^{-1}]_{ji} = \frac{D_{i-1} D_{N-j}}{D_N} \;\; ; G_{ij} = x^{i-j-1} \;\; \text{if} \;\; |x| >> 2 \qquad \text{(A.4)}$$

$$\frac{x}{2} = \frac{z - E_{Dot}}{2t_{Dot}} = \begin{cases} \cos\Phi & \text{if } |x| < 2 \\ \pm 1 & \text{if } |x| = 2 \\ \cosh\Phi & \text{if } |x| > 2 \end{cases} \qquad \text{(A.5)}$$

$$D(N,\alpha,\beta) = \frac{\sin(N+1)\Phi}{\sin\Phi} - (\alpha+\beta)\frac{\sin N\Phi}{\sin\Phi} + \alpha\beta\,\frac{\sin(N-1)\Phi}{\sin\Phi} \quad \text{(A.6)}$$

Alternatively, one can always assume $x = 2\cos\Phi$ and allow $\Phi$ to be imaginary if $z$ is outside the band of the dot. After substitution in (A.6), one finds that:

$$D(N,\alpha,\beta)=\begin{cases}\dfrac{\sin(N+1)\Phi}{\sin\Phi}-(\alpha+\beta)\dfrac{\sin N\Phi}{\sin\Phi}+\alpha\beta\dfrac{\sin(N-1)\Phi}{\sin\Phi}\\ N[(1-\alpha)(1-\beta)]+1-\alpha\beta\\ \dfrac{\sinh(N+1)\Phi}{\sinh\Phi}-(\alpha+\beta)\dfrac{\sinh N\Phi}{\sinh\Phi}+\alpha\beta\dfrac{\sinh(N-1)\Phi}{\sinh\Phi}\end{cases} \quad (A.7)$$

# Acknowledgment

We thank of A . T . Rezakhani for edit this paper.